# Ordered addition of two Lorentz boosts through spatial and space-time rotations

Chandru Iyer[1]
G. M. Prabhu[2]

[1]Techink Industries, C-42, phase-II, Noida, India 201305
[2]Department of Computer Science, Iowa State University, Ames, IA 50011, USA
Contact E-mail: prabhu@cs.iastate.edu

**Abstract**
The ordered addition of two Lorentz boosts is normally shown to result in a boost by utilizing concepts from group theory and non-Euclidian geometry. We present a method for achieving this addition by performing a sequence of spatial rotations and uni-dimensional Lorentz transformations. The method is first developed for two-dimensional space and it is then extended to three-dimensional space by utilizing the commutative property of the rotation of the *y-z* plane and a boost along the *x*-axis. The method employs only matrix multiplication and certain invariant quantities that are natural consequences of spatial rotations and Lorentz transformations. The combining of two boosts in different directions into a single boost cannot be expected *a priori* because we show that the converse of this statement is not true. That is, two rotations interspersed with a boost cannot always be reduced to a single rotation preceded and followed by boosts.

## 1. Introduction

The velocity addition formula for co-linear velocities (that is, velocities in the same direction) is easily established in the theory of Special Relativity by considering the transformation of the coordinates of a moving object P as observed by two inertial frames K and K' which themselves are in relative motion [1, 2]. This leads to the well known formula of co-linear velocity addition $(u+v)/(1 + uv/c^2)$. Møller extended this approach to develop the velocity addition formula for combining two velocities in different directions in planar motion [3]. It is also well known that the addition of two boosts in planar motion involves an additional rotation of the line of motion. This additional rotation is called Thomas Precession. While the co-linear addition of velocities is commutative, the non-colinear addition (either planar or in 3d-space) is a non-commutative operation and the order in which the velocities are added affects the resultant velocity [4].

In this paper we develop a method for adding or combining two boosts, which are arbitrarily oriented in 3-dimensional space. It should be noted that the orientation of the second boost is well defined only from the point of view of an observer co-moving with the first boost. This makes it necessary to perform the second boost from an inertial frame K', co-moving with the first boost and in the context of the synchronicity that is unique to the inertial frame K'. These considerations make the 'addition' operation of two boosts non-commutative [4].

The combining of boosts in the same direction is normally shown with rigor using the transformation equations [1, 2]. For combining boosts in different directions, techniques such as group theory and gyro-vectors have been suggested in the literature [5]. These techniques are abstract and elaborate. We propose a simple constructive method using only matrix multiplication and a few invariant equations that are natural consequences of spatial rotations and Lorentz transformations. The combining of boosts interspersed with spatial rotation into a single boost is a result that cannot be expected *a priori* because we show that its converse, which is a logically similar statement, namely, "spatial rotations interspersed with boosts become a single rotation preceded and followed by boosts," is false.

We first develop a method for the ordered addition of two arbitrarily oriented planar boosts. We extend this to 3-dimensions by noting that even though a spatial rotation and a Lorentz boost are normally not commutative, in the special case of the rotation of the *y-z* plane and a boost along the *x*-axis, these two operations remain commutative. In other words, a rotation of the plane perpendicular to a boost can be performed equivalently either before or after the transformation associated with the boost. The pedagogical contribution of this paper is the development of a constructive method for the addition of planar boosts and its extension for the addition of three-dimensional boosts.

## 2. Addition of Planar boosts

### 2.1 The LRL transformation: Two boosts in different directions in a plane

In this section we derive the transformation matrix for a conventional Lorentz ($L_{xu}$) followed by a planar rotation of the *x-y* plane ($R_\theta$) and then followed by another conventional Lorentz ($L_{xv}$). For clarity, we may visualize four inertial reference frames K, L, M, and N in two-dimensional space. Frames K and L have both their coordinate axes aligned and L is moving at a velocity *u* along the *x*-axis as observed by K. The inertial frame L has another coordinate reference frame M, where the axes of M are rotated by an angle θ counterclockwise with respect to L. Frames M and N have both their coordinate axes aligned and N is moving at a velocity *v* along the *x*-axis as observed by M. The event coordinate transformation from K to N is given by matrix **G** which is equal to the matrix product $\mathbf{L_{xv} \, R_\theta \, L_{xu}}$. The matrices $R_\theta$ and $L_{xv}$ are as specified in equations (1) and (2) respectively, using the notation described in [6].

$$\begin{pmatrix} x' \\ y' \\ t' \end{pmatrix} = \begin{pmatrix} \cos\theta & \sin\theta & 0 \\ -\sin\theta & \cos\theta & 0 \\ 0 & 0 & 1 \end{pmatrix} \begin{pmatrix} x \\ y \\ t \end{pmatrix} \qquad (1)$$

**R$_\theta$** denotes the above transformation, indicating spatial rotation anticlockwise by an angle θ.

$$\begin{pmatrix} x' \\ y' \\ t' \end{pmatrix} = \begin{pmatrix} \gamma & 0 & -v\gamma \\ 0 & 1 & 0 \\ -v\gamma/c^2 & 0 & \gamma \end{pmatrix} \begin{pmatrix} x \\ y \\ t \end{pmatrix} \qquad (2)$$

**L$_{xv}$** denotes a Lorentz transformation of magnitude *v* along the *x*-axis.

The matrix $L_{xu}$ is the same as the one in equation (2) with velocity *u* substituted for *v*. The inverse of this operation is given by $\mathbf{H = L_{x(-u)} \, R_{(-\theta)} \, L_{x(-v)}}$. Matrices G and H are obtained reciprocally from each other by replacing *u* by *–v*, *v* by *–u* and θ by *–θ*. Matrices G and H are inverse of each other as the processes associated with them are inverse of each other. The elements of matrices G and H are obtained by matrix multiplication of the corresponding matrices specified in equations (1) and (2) with appropriate values for $v$ and $\theta$. Matrices G and H turn out as follows.

$$G = \mathbf{L_{xv}} \, \mathbf{R_\theta} \, \mathbf{L_{xu}} = \begin{pmatrix} \gamma_v & 0 & -v\gamma_v \\ 0 & 1 & 0 \\ -v\gamma_v/c^2 & 0 & \gamma_v \end{pmatrix} \begin{pmatrix} \cos\theta & \sin\theta & 0 \\ -\sin\theta & \cos\theta & 0 \\ 0 & 0 & 1 \end{pmatrix} \begin{pmatrix} \gamma_u & 0 & -u\gamma_u \\ 0 & 1 & 0 \\ -u\gamma_u/c^2 & 0 & \gamma_u \end{pmatrix}$$

$$= \begin{pmatrix} \left(\cos\theta + \dfrac{uv}{c^2}\right)\gamma_u\gamma_v & \gamma_v\sin\theta & -\gamma_v\gamma_u(v+u\cos\theta) \\ -\gamma_u\sin\theta & \cos\theta & u\gamma_u\sin\theta \\ \dfrac{-\gamma_u\gamma_v}{c^2}(u+v\cos\theta) & \dfrac{-v\gamma_v\sin\theta}{c^2} & \gamma_u\gamma_v\left(1+\dfrac{uv\cos\theta}{c^2}\right) \end{pmatrix} \quad (3)$$

$$H = \mathbf{L_{x(-u)}} \, \mathbf{R_{(-\theta)}} \, \mathbf{L_{x(-v)}} = \begin{pmatrix} \left(\cos\theta + \dfrac{uv}{c^2}\right)\gamma_u\gamma_v & -\gamma_u\sin\theta & \gamma_v\gamma_u(u+v\cos\theta) \\ \gamma_v\sin\theta & \cos\theta & v\gamma_v\sin\theta \\ \dfrac{\gamma_u\gamma_v}{c^2}(v+u\cos\theta) & \dfrac{-u\gamma_u\sin\theta}{c^2} & \gamma_u\gamma_v\left(1+\dfrac{uv\cos\theta}{c^2}\right) \end{pmatrix} \quad (4)$$

where $\gamma_u = \dfrac{1}{\sqrt{1-u^2/c^2}}$ and $\gamma_v = \dfrac{1}{\sqrt{1-v^2/c^2}}$.

The matrix G can be visualized as the addition of $u$ and $v$ in that order with an orientation shift of $\theta$ in between. Similarly, the matrix H can be visualized as the addition of $-v$ and $-u$ in that order with an orientation shift of $-\theta$ in between.

It can be verified that GH = HG = I, the Identity matrix, by algebraic and trigonometric simplifications. It may be noted that the matrices G and H were independently obtained by considering the transformations between K, L, M, and N in that order (for G) and N, M, L, and K in that order (for H). As expected, mathematically, the matrices G and H turn out to be the inverse of each other. Further, we observe that the diagonal elements of G and H are identical.
We can use matrices G and H to transform coordinates between the two frames K and N as follows.

$$\begin{pmatrix} x' \\ y' \\ t' \end{pmatrix} = G \begin{pmatrix} x \\ y \\ t \end{pmatrix} \quad \text{---- (5a);} \qquad \begin{pmatrix} x \\ y \\ t \end{pmatrix} = H \begin{pmatrix} x' \\ y' \\ t' \end{pmatrix} \quad \text{---- (5b)}$$

For the origin of N, by setting $x' = 0$ and $y' = 0$ in equation (5b) one can derive the velocity addition formula as described in standard text books such as [3].

## 2.2 The RLR transformation: A planar boost preceded and followed by planar rotations

In this section we derive the transformation matrix for RLR, that is, a spatial rotation $\phi$ followed by a conventional Lorentz and then followed by another spatial rotation $\alpha$ as described in [6].

$$\mathbf{R_\alpha\ L_{(xw)}\ R_\phi} = \begin{pmatrix} \cos\alpha & \sin\alpha & 0 \\ -\sin\alpha & \cos\alpha & 0 \\ 0 & 0 & 1 \end{pmatrix} \begin{pmatrix} \gamma_w & 0 & -w\gamma_w \\ 0 & 1 & 0 \\ -w\gamma_w/c^2 & 0 & \gamma_w \end{pmatrix} \begin{pmatrix} \cos\phi & \sin\phi & 0 \\ -\sin\phi & \cos\phi & 0 \\ 0 & 0 & 1 \end{pmatrix}$$

We denote this transformation matrix as D whose elements are as follows:

$$D = \begin{pmatrix} \gamma_w \cos\alpha\cos\phi - \sin\alpha\sin\phi & \gamma_w \cos\alpha\sin\phi + \sin\alpha\cos\phi & -w\gamma_w\cos\alpha \\ -\gamma_w \sin\alpha\cos\phi - \cos\alpha\sin\phi & -\gamma_w \sin\alpha\sin\phi + \cos\alpha\cos\phi & w\gamma_w\sin\alpha \\ \dfrac{-w\gamma_w\cos\phi}{c^2} & \dfrac{-w\gamma_w\sin\phi}{c^2} & \gamma_w \end{pmatrix} \quad (6)$$

where $\gamma_w = \dfrac{1}{\sqrt{1-w^2/c^2}}$.

Similar to the case with matrices G and H, the inverse of matrix D, corresponding to the transformation $\mathbf{R_{(-\phi)}\ L_{x(-w)}\ R_{(-\alpha)}}$ is denoted as matrix E with the following elements:

$$E = \begin{pmatrix} \gamma_w \cos\phi\cos\alpha - \sin\phi\sin\alpha & -\gamma_w \cos\phi\sin\alpha - \sin\phi\cos\alpha & w\gamma_w\cos\phi \\ \gamma_w \sin\phi\cos\alpha + \cos\phi\sin\alpha & -\gamma_w \sin\phi\sin\alpha + \cos\phi\cos\alpha & w\gamma_w\sin\phi \\ \dfrac{w\gamma_w\cos\alpha}{c^2} & \dfrac{-w\gamma_w\sin\alpha}{c^2} & \gamma_w \end{pmatrix} \quad (7)$$

## 3. Comparison of the LRL and RLR Planar Transformations

### 3.1 Some characteristics of the LRL and RLR transformation matrices

The LRL matrices G and H described in equations (3) and (4) have a resultant velocity which can be evaluated by considering the motion of the origin of N $(x' = 0;\ y' = 0)$ as

$$\mathbf{u + v} = \dfrac{\sqrt{u^2 + v^2 + 2uv\cos\theta - u^2 v^2 \dfrac{\sin^2\theta}{c^2}}}{\dfrac{uv\cos\theta}{c^2} + 1}. \quad (8)$$

This result is also derived in [3] by a different approach. Although the expression in equation (8) appears to be commutative, the commutativity is restricted only to the magnitude of the velocity [4]. The directions of $\mathbf{u + v}$ and $\mathbf{v + u}$ are different and therefore the operation is not commutative.

The last element (3, 3) of both G and H is equal to the value of γ corresponding to the resultant velocity given in equation (8). The RLR matrices D and E described in equations (6) and (7) also have $\gamma_w$ as the (3, 3) element. Furthermore, all these matrices have 3 parameters. In the case of the LRL transformation, the 3 parameters are $u$, $\theta$, and $v$. In the case of the RLR transformation, the 3 parameters are $\phi$, $w$, and $\alpha$. The matrices themselves have nine elements constructed from the 3 parameters.

Matrices G and D have to obey the following invariance for all space-time points $(x, y, t)$ and $(x', y', t')$:

$$x^2 + y^2 - c^2 t^2 = x'^2 + y'^2 - c^2 t'^2. \tag{9}$$

Equation (9) places six constraints on the elements of G, H, D, and E. These constraints are obtained by equating the coefficients of $x^2$, $y^2$, $t^2$ on both sides and setting the coefficients of $xy$, $yt$, $tx$ as zero after expanding the right hand side of equation (9).

In particular, we will be utilizing two of these constraints, equations (10a) and (10b) below, for converting a G matrix to a D matrix.

$$g_{33}^2 - \frac{(g_{13}^2 + g_{23}^2)}{c^2} = 1 \,;\, d_{33}^2 - \frac{(d_{13}^2 + d_{23}^2)}{c^2} = 1 \,; \tag{10a}$$

$$g_{33}^2 - c^2(g_{31}^2 + g_{32}^2) = 1 \,;\, d_{33}^2 - c^2(d_{31}^2 + d_{32}^2) = 1 \,; \tag{10b}$$

We show that given $u$, $\theta$, and $v$, we can always generate an equivalent set of $\phi$, $w$, and $\alpha$ such that G $(u, \theta, v)$ is identical to D $(\phi, w, \alpha)$; this procedure physically means the addition of the two planar boosts **u** and **v** resulting in **w.**

However, we note that the reverse of this, i.e., conversion from D to G is possible only sometimes because in matrices G and D, the range of $g_{22}$ is $-1$ to $+1$ and the range of $d_{22}$ is $-\gamma$ to $+\gamma$.

### 3.2 Procedure for converting LRL (G–Matrix) to RLR (D–matrix)

In general, any LRL transformation (G matrix) created by the three parameters $u$, $\theta$, and $v$, can be converted to an RLR transformation as follows.

We first set $g_{33} = \gamma$ and obtain $w = \sqrt{1 - (1/\gamma^2)}$ (it can be shown that $g_{33} \geq 1$ whenever $|u| \leq c$ and $|v| \leq c$).

We then express $g_{13}$ and $g_{23}$ as follows: $g_{13} = -w\gamma p$ and $g_{23} = w\gamma q$.

From the constraints dictated by equation (10a), it can be seen that $p^2 + q^2 = 1$. Therefore $p$ and $q$ can be expressed as $\cos\alpha$ and $\sin\alpha$. Thus, by considering the last column of matrices G and D, we extract the parameter α for a given G matrix as

$$\alpha = \cos^{-1}\left(\frac{-g_{13}}{g_{33} * w}\right).$$

A similar procedure applied to the third row of G and D, along with equation (10b), gives us the parameter $\phi$ as

$$\phi = \cos^{-1}\left(\frac{-g_{31}}{g_{33} * (w/c^2)}\right).$$

The quadrant of the angles $\phi$ and α may be chosen appropriately by inspecting the sign of elements $g_{32}$ ($d_{32}$) and $g_{23}$ ($d_{23}$) respectively. So by considering the third row and third column of a G matrix with parameters $u$, $\theta$, and $v$, we can always extract parameters $\phi$, $w$, and α, of an equivalent D matrix; the other four elements of G and D, namely elements (1, 1), (1, 2), (2, 1), and (2, 2) are equal because of the constraints imposed by equation (9) on both G and D. The D matrix, when $|d_{22}| > 1$, cannot be converted to an equivalent G matrix. It is to be noted that the range of $d_{22}$ is $-\gamma_w$ to $+\gamma_w$ and the range of $g_{22}$ is $-1$ to $+1$. Matrix D can be converted to an equivalent matrix G by a similar procedure, considering the second row and second column of D and G, only when $|d_{22}| \leq 1$. On the other hand, matrix G can always be converted to an equivalent matrix D.

Thus two planar boosts, which are arbitrarily oriented (represented by a G matrix), can always be converted to a single boost preceded and followed by a planar rotation (represented by a D matrix). The converse of this statement, that is, two planar rotations interspersed with a Lorentz boost (RLR) combine to become a planar rotation preceded and followed by boosts (LRL) is not (always) true. Therefore, it is difficult to expect the first result intuitively. We have provided a constructive proof using the transformations directly.

For planar motion we have also developed an Excel spreadsheet that automatically converts an LRL transformation to a RLR transformation and can be downloaded from [7].

**4. Combining boosts in 3-dimensional space**

In this section we extend our method to 3-dimensional space and show that two boosts in arbitrary directions in a 3-dimensional space add up to a single boost. Although this fact has been established by group theory and other mathematical techniques, our method is constructive.

**4.1 Representation of a boost in an arbitrary direction in 3-d space**

A boost with a unit vector along its direction as represented in equation (11)

$$\cos\theta \, \mathbf{i} + \cos\phi \, \sin\theta \, \mathbf{j} + \sin\phi \, \sin\theta \, \mathbf{k} \tag{11}$$

can be visualized as subtending an angle $\theta$ with the *x*-axis and *(90 – θ)* with the *y-z* plane. To clarify further, we need to visualize the plane in which the *x*-axis and the unit vector along the boost under consideration are situated. This plane intersects the *y-z* plane along a line L. This line L, which is on the *y-z* plane and perpendicular to the *x*-axis subtends an angle *(90 – θ)* with the unit vector along the boost. Furthermore, L subtends an angle $\phi$ with the *y*-axis. Thus we arrive at equation (11) above for the unit vector along the boost in terms of the unit vectors along the coordinate axes namely **i**, **j**, and **k**.

**4.2 Appropriate rotations for aligning the boost with the *x*-axis**

By performing the following spatial rotations, we can align the *x*-axis along the boost.

- (1) Rotate the *y-z* plane by an angle $\phi$
- (2) Rotate the *x-y* plane by an angle $\theta$

Although the alignment of the *x*-axis and the line of a boost can be achieved by the more obvious combination of spatial rotations $\mathbf{R}_{xz}\mathbf{R}_{xy}$, we choose the alternative $\mathbf{R}_{xy}\mathbf{R}_{yz}$ by hindsight, to utilize the commutative property of $\mathbf{R}_{yz}$ and a boost along the *x*-axis. In the chosen $\mathbf{R}_{xy}\mathbf{R}_{yz}$ combination, the first rotation $\mathbf{R}_{yz}$ rotates the *yz* plane such that the *y*-axis is positioned in such a way that the *x*-axis, the line of the boost, and the *y*-axis, all lie on a plane. This enables the next rotation $\mathbf{R}_{xy}$ which rotates the (new) *xy* plane to align the *x-a*xis with the line of the boost.

The two operations are to be performed in the order specified. The first operation is represented by the matrix:

$$\mathbf{R}_{yz(\varphi)} = \begin{pmatrix} 1 & 0 & 0 & 0 \\ 0 & \cos\phi & \sin\phi & 0 \\ 0 & -\sin\phi & \cos\phi & 0 \\ 0 & 0 & 0 & 1 \end{pmatrix}$$

Similarly the second operation is represented by the matrix:

$$\mathbf{R}_{xy(\theta)} = \begin{pmatrix} \cos\theta & \sin\theta & 0 & 0 \\ -\sin\theta & \cos\theta & 0 & 0 \\ 0 & 0 & 1 & 0 \\ 0 & 0 & 0 & 1 \end{pmatrix}$$

The tip of the unit vector along the boost has the following coordinates:

$$\begin{pmatrix} x \\ y \\ z \\ t \end{pmatrix} = \begin{pmatrix} \cos\theta \\ \cos\phi\sin\theta \\ \sin\phi\sin\theta \\ t \end{pmatrix},$$

where *t* is any instant in the inertial frame K. It is easy to verify that

$$\mathbf{R}_{xy(\theta)}\,\mathbf{R}_{yz(\varphi)} \begin{pmatrix} \cos\theta \\ \cos\phi \sin\theta \\ \sin\phi \sin\theta \\ t \end{pmatrix} = \begin{pmatrix} 1 \\ 0 \\ 0 \\ t \end{pmatrix}.$$

Thus we find that the two operations (done in the order indicated) $\mathbf{R}_{xy(\theta)}\,\mathbf{R}_{yz(\varphi)}$ align the x-axis with the direction of the boost as given in equation (11).

**4.3 Addition of two boosts in 3-d space**

For simplicity we assume that the x-axis is aligned with the first boost of magnitude u. To transport to the inertial frame M moving at velocity u along the x-axis with respect to inertial frame K, we perform the conventional Lorentz transformation $\mathbf{L}_{xu}$. The second boost of magnitude v may be in an arbitrary direction. So we first align the x-axis of M with the second boost as indicated in Section 4.2 and then perform a conventional Lorentz of magnitude v to get

T = $\mathbf{L}_{xv}\,\mathbf{R}_{xy(\theta)}\,\mathbf{R}_{yz(\varphi)}\,\mathbf{L}_{xu}$ (12)

The above transformation T takes us to an inertial frame K' moving with a velocity v with respect to frame M along the direction of the second boost.

It can be easily shown by simple matrix multiplication (which in general is not commutative) that $\mathbf{R}_{yz(\varphi)}\,\mathbf{L}_{xu} = \mathbf{L}_{xu}\,\mathbf{R}_{yz(\varphi)}$.

The physical justification of this property of commutativity is because the transformation $\mathbf{L}_{xu}$ between x and t and the transformation $\mathbf{R}_{yz(\varphi)}$ between y and z are mutually exclusive and have no common parameter. Therefore, equation (12) can be rewritten as

T = $\mathbf{L}_{xv}\,\mathbf{R}_{xy(\theta)}\,\mathbf{L}_{xu}\,\mathbf{R}_{yz(\varphi)}$. (13)

The product $\mathbf{L}_{xv}\,\mathbf{R}_{xy(\theta)}\,\mathbf{L}_{xu}$ in equation (13) is on the x-y plane and we can use the procedure described in Section 3.2 to convert this LRL to RLR – that is, to $\mathbf{R}_{xy(A)}\,\mathbf{L}_{xw}\,\mathbf{R}_{xy(B)}$. Thus we obtain

T = $\mathbf{R}_{xy(A)}\,\mathbf{L}_{xw}\,\mathbf{R}_{xy(B)}\,\mathbf{R}_{yz(\varphi)}$. (14)

Equation (14) proves that two boosts of u and v in different directions in 3-dimensional space result in a single boost of w with appropriate spatial rotations preceding and following the boost. The inverse of such a boost is also a boost. Two inertial frames that are connected through a series of inertial frames related by boosts, have a single boost that defines the transformation between them, as illustrated below.

$\mathbf{I_3} \leftarrow B_{13} \text{———} \mathbf{I_1} \text{———} B_{12} \rightarrow \mathbf{I_2}$

The transformation between $\mathbf{I_2}$ to $\mathbf{I_3}$ is given by $B_{12}^{-1}B_{13}$ which is equal to $B_{21}B_{13}$. As long as $B_{12}$ and $B_{13}$ are combinations of spatial rotations and uni-dimensional Lorentz transformations, the product $B_{21}B_{13}$ will also be a transformation of the same type.

## 5. Conclusions

We developed a formulation for combining two arbitrarily oriented boosts on a plane. We then extended it to 3-dimensional motion by observing the property of commutativity of the rotation of the plane perpendicular to a boost and the transformation associated with that boost. Thus the combining of two boosts arbitrarily oriented in three spatial dimensions can be reduced to combining two boosts on a plane, which is accomplished by converting an LRL matrix to a RLR matrix.

Students can actually download the Excel spreadsheet in [7] for adding two planar boosts of their choice. We have found this spreadsheet useful in understanding the extra rotation known as Thomas Precession. Our extension of the planar method to three-dimensions (by noting the commutativity of the rotation of the plane perpendicular to the boost and the transformation associated with the boost) completes our objective of the ordered addition of two arbitrary boosts in three-dimensional space by a constructive methodology. It further emphasizes the fact that two three-dimensional boosts can be reduced to planar boosts by an appropriate choice of the coordinate axes.

The one-dimensional addition of boosts has a constructive background. It uses the transformation equations directly. In a similar manner, we have 'added' multi-dimensional boosts by the stepwise application of the transformations associated with spatial rotations and Lorentz boosts. The combining of boosts interspersed with spatial rotations to a single boost cannot be expected *a priori* because we have shown that its converse is not always true. That is, rotations interspersed with boosts cannot (always) be reduced to a single rotation preceded and followed by boosts. Our method proves that boosts in different directions always combine to become a single boost, by a constructive methodology without using techniques such as group theory, non-Euclidian geometry, or gyro-vectors. Neither does our method assume the existence of a four-dimensional continuum. One conclusion that can be drawn is that the nature of the associated transformation permits boosts in different directions to become a single boost, irrespective of whether a four-dimensional continuum exists or not.